\documentclass[doublecol,figures]{epl2}
\usepackage{color,amssymb}
\newcommand{\Heff}{\mathcal{H}_{\mathrm{eff}}}
\newcommand{\aver}[1]{\left\langle #1 \right\rangle}

\title{Resonance width distribution in RMT: Weak coupling regime \\ beyond Porter-Thomas}

\author{Yan V. Fyodorov$^1$ \and Dmitry V. Savin$^{2,3}$}

\institute{
 $^1$ Queen Mary University of London, School of Mathematical Sciences, London E1 4NS, United Kingdom \\
 $^2$ Department of Mathematics, Brunel University London, Uxbridge, UB8 3PH, United Kingdom \\
 $^3$ Max Planck Institute for the Physics of Complex Systems, 01187 Dresden, Germany
}
\pacs{05.45.Mt}{Quantum chaos; semiclassical methods}
\pacs{03.65.Nk}{Scattering theory}
\pacs{05.60.Gg}{Quantum transport}

\abstract{We employ the random matrix theory (RMT) framework to revisit the distribution of resonance widths in quantum chaotic systems weakly coupled to the continuum via a finite number $M$ of open channels.  In contrast to the standard first-order perturbation theory treatment we do not {\it a priory} assume the resonance widths being small compared to the mean level spacing.  We show that to the leading order in weak coupling the perturbative $\chi^2_M$ distribution of the resonance widths (in particular, the Porter-Thomas distribution at $M=1$) should be corrected by a factor related to a certain average of the ratio of square roots of the characteristic polynomial (`spectral determinant') of the underlying RMT Hamiltonian. A simple single-channel expression is obtained that properly approximates the width distribution also at large resonance overlap, where the Porter-Thomas result is no longer applicable.}

\begin{document}
\maketitle

\textbf{Introduction.}-- Scattering of both classical and quantum waves in systems with chaotic intrinsic dynamics is characterised by universal statistical properties \cite{Stoeckmann,kuhl05a,grad14}. Those can be understood by studying the properties of quasi-stationary states in an open system formed at the intermediate stage of the scattering process \cite{soko89,fyod97,fyod05,mitc10}. Such states are associated with the resonances which correspond to the complex poles $\mathcal{E}_n=E_n-\frac{i}{2}\Gamma_n$ of the $S$ matrix. The resonance widths $\Gamma_n>0$ bring fundamentally new features to the open system in comparison with its closed counterpart. In particular, width fluctuations govern the decay law \cite{savi97,ditt00} and give rise to nonorthogonal modes \cite{schom00,savi06b} leading to enhanced sensitivity to perturbations \cite{fyod12b}. Various aspects of lifetime and width statistics are a subject of intensive research, both theoretically and experimentally, with recent studies motivated by practical applications including optical microresonators \cite{schom09}, superconductor superlattices \cite{glue02}, many-body fermionic systems \cite{cela11}, microwave billiards \cite{poli10,weich14}, and dissipative quantum maps \cite{keat06,nova13a}, as well as by long-standing interest in superradiance-like ``resonance trapping'' phenomena  \cite{soko89,pers00,cela11,auer11,mulh15}. Recent advances in experimental techniques have made it possible to test many of theoretical predictions with unprecedented accuracy \cite{kuhl08,diet10,koeh10,difa12,liu14,bark13,kuhl13,gros14}, giving further impetus for the theory development.

The properties of resonance states depend crucially on the way in which the system is coupled to the continuum. In the case of the big (``classical'') opening, the wave escape follows ballistic trajectories and is dominated by system-specific characteristics of chaotic dynamics. The quasibound states then have the support in the associated classical repeller \cite{lu03}, i.e. the (fractal) set of rays that remain trapped in the scattering region, and thus live long enough to develop random-wave properties \cite{schom04}. The cloud of resonances is formed such that it is typically separated by a finite gap from the real axis, with the gap width being of the order of the classical decay rate  \cite{gasp89i} (see also \cite{haak92,lehm95a}). Finding resonances inside that gap is a rare event, see \cite{nova13a} for a recent review and further discussion. Inside the cloud the resonance distribution  mainly follows the so-called fractal Weyl law \cite{lu03,schom04}, which is determined by the fractal dimension of the chaotic repeller. This and related characteristics have attracted considerable attention recently both in physics \cite{nova13a,altm13} and mathematical literature \cite{nonn14}.

In many realistic situations, however, the opening is small (``quantum'') in the sense that it does not scale with the system size and supports at a given energy only a finite number $M$ of scattering channels. Representative examples are microwave cavities \cite{Stoeckmann} (where $M$ is just the number of the attached antennas) and quantum dots \cite{alha00} (then $M$ is fixed by transverse quantisation for the modes propagating in the leads). Under such a condition, the waves remain in the system long enough to fully explore the whole available phase-space and to develop strong diffraction and interference effects, leading to decay rates $\Gamma_n$ being of the order of the mean level spacing $\Delta$. As a result, the scattering matrix and related observables exhibit statistical fluctuations on the scale $\sim\Delta$, which are universal, i.e. system-independent, provided that the appropriate natural units have been used \cite{guhr98}. The only control parameters are then the so-called transmission coefficients $T_c=1-|\overline{S}_{cc}|^2$, with $\overline{S}$ being the average `optical' $S$ matrix. They quantify the system openness, so that $T_c=1$ ($T_c\to0$) stands for a perfectly open (almost closed) channel $c=1,\ldots,M$.

The case of the weakly open system, all $T_c\ll1$, is of particular interest, since it is most easily realised experimentally. The resonances are then well separated, so one can use perturbation theory \cite{port56} to express the widths $\Gamma_n=\sum_c|V_n^c|^2\ll\Delta$ in terms of the partial amplitudes $V_n^c$ to decay into channel $c$. The latter have Gaussian statistics inherited from chaotic wavefunctions of the closed system. For equivalent channels, the distribution of the resonance widths $\kappa_n=\Gamma_n/\overline\Gamma$  measured in units of the mean partial width $\aver{|V_n^c|^2}\equiv\overline{\Gamma}$ is readily found to be given by the $\chi^2$ distribution with $M\beta$ degrees of freedom,
\begin{equation}\label{PTdis}
 \chi^2_{M\beta}(\kappa) =
   \frac{(\beta/2)^{M\beta/2} }{ \Gamma(M\beta/2)} \kappa^{M\beta/2-1} e^{-\beta\kappa/2}\,,
\end{equation}
where $\beta=1$ ($\beta=2$) indicates the systems with preserved (broken) time-reversal symmetry. Distribution (\ref{PTdis}) has the mean value $\aver{\kappa}=M$ and variance $\mathrm{var}(\kappa)=\frac{2}{M\beta}\aver{\kappa}^2$, showing that typical resonances are indeed very narrow for $\overline\Gamma/\Delta\ll1$. The regime of overlapping resonances is clearly beyond the reach of (\ref{PTdis}). In particular, it fails to comply with the exact Moldauer-Simonius sum rule \cite{mold68,simo74b} for the average width $\aver{\Gamma}_\mathrm{MS} = -\frac{\Delta}{2\pi}\sum_c\ln(1-T_c)$.  Note that it has a log divergence as $T_c\to1$ related to the emerging universal powerlaw far tail $\sim\Gamma^{-2}$\cite{fyod96b} of the width distribution in such a regime. Describing a crossover to such behavior requires going beyond the perturbative scheme used to derive (\ref{PTdis}) and will be discussed below.

For a single open channel, distribution (\ref{PTdis}) at $\beta=1$ was first obtained in the classical work by Porter and Thomas\cite{port56} and bears their name (PTD). It was found to agree favorably with early experiments on elastic scattering of slow neutrons off heavy nuclei. In fact, for a long time that was considered among most firmly established manifestations of chaos in compound nuclear scattering \cite{mitc10}. However, recent experiments with improved accuracy \cite{koeh10} found evidence of significant deviations from the PTD. Various physical mechanisms possibly contributing to the discrepancy were discussed in the literature\cite{weid10,voly11,cela11}, and motivated a few proposals towards its resolution. In particular, one of the proposals \cite{shched12} suggested a semi-heuristic way of going beyond the simple first-order perturbation theory for $M=1$ and $\beta=1$, in order to get access to the improved description of those (rare) resonances whose reduced widths $\Gamma_n/\Delta\sim1$ despite the weak coupling limit.

The main goal of our letter is to demonstrate that the existing body of knowledge already available (though not explicitly worked out) in the literature about the $\beta=1$ case allows one to derive the required expression in a fully controllable way. For $M=1$ we obtain a simple explicit expression (\ref{singlechannel}) that properly approximates the width distribution even at  $\Gamma_n\gg\Delta$. Comparing our result to the heuristic proposal \cite{shched12} is instructive, and reveals some interesting similarities and important differences. We also reconsider this case from a perspective which is close (but not identical) to the spirit of \cite{shched12}. This allows us to give a RMT meaning to the correction factor that arises.

\textbf{RMT formulation.}\,-- As is well known \cite{soko89,fyod97} (see also \cite{mitc10,fyod11ox} for recent reviews), the non-perturbative approach to address universal statistics of $N$ resonance states coupled to $M$ channels is based on applying RMT to the  effective non-Hermitian Hamiltonian of the open system
\begin{equation}\label{Heff}
  \Heff^{(N)} = H - \frac{i}{2}VV^\dag\,.
\end{equation}
Here $H$ is the Hermitian $N{\times}N$ matrix that corresponds to the Hamiltonian of the closed quantum chaotic system, whereas the rectangular $N{\times}M$ matrix $V$ consists of the coupling amplitudes between $N$ energy levels of $H$ and $M$ open scattering channels. The poles of the $S$ matrix are just given by $N$ complex eigenvalues of $\Heff$ which therefore coincide with the resonances $\mathcal{E}_n$. Their universal fluctuations occur in the RMT limit $N\gg1$ (but $M$ kept finite) at the local scale $\Delta\sim\frac{1}{N}$, which can be modelled by random $H$ drawn from Gaussian orthogonal (GOE, $\beta=1$) or unitary (GUE, $\beta=2$) ensemble, depending on the symmetry under time-reversal. Without loss of generality, one can restrict the consideration to the spectrum center, where the mean level spacing $\Delta=\lambda\pi/N$ and $2\lambda$ is the semicircle radius ($\Delta$ needs to be rescaled if $E\neq 0$). As long as $M\ll N$, the results do not depend on statistical assumptions on the coupling amplitudes \cite{lehm95a}, which are usually taken either as $M$ random\cite{soko89} or fixed \cite{verb85} $N$-component vectors characterised by the auxiliary coupling constants $\gamma_c=\pi\|V^c\|^2/(2N\Delta)$. The latter enter final expressions only through the transmission coefficients $T_c=4\gamma_c/(1+\gamma_c)^2\leq1$, so that $\gamma_c\ll1$ ($\gamma_c=1$) corresponds to weak (perfect) coupling. It is also convenient to use the renormalized coupling strengths defined as
\begin{equation}\label{g_c}
 g_c \equiv \frac{2}{T_c}-1 = \frac{1}{2}\left(\gamma_c+\frac{1}{\gamma_c}\right)\geq1\,.
\end{equation}

The standard treatment of the regime of weak coupling to the continuum amounts to considering the anti-Hermitian part of $\Heff$ as a perturbation of the Hermitian part $H$. To the leading order in $\gamma_c\ll1$, the resonance widths are given by $\Gamma_n = \langle{n}|VV^\dag|n\rangle=\sum_c|V_n^c|^2$, where $|n\rangle$ stands for the eigenvector corresponding to the $n$-th eigenvalue (energy level) of $H$. A rotation that diagonalizes random $H$ ensures that in the limit $N\gg1$ the components $V_n^c$ become independent Gaussian distributed variables. For equivalent channels this readily gives (\ref{PTdis}), with $\overline{\Gamma}=2\gamma\Delta/\pi\approx T\Delta/2\pi$. (The generalisation for nonequivalent and correlated weak channels is also known \cite{alha95}.)

A non-pertubative approach to the resonance distribution is based on making use of an electrostatic analogue \cite{soko89,lehm95a,fyod97} treating the $S$ matrix poles as two-dimensional charges and employing the powerful supersymmetry technique \cite{verb85} to perform the statistical average. Let us stress that in such a way the distribution of the resonance widths was already derived for any number of arbitrary open channels first at $\beta=2$ \cite{fyod97,fyod96b} and then at $\beta=1$ \cite{somm99}. Unfortunately, in such generality the exact $\beta=1$ expression takes a rather complicated form of a three-fold integral and is not very transparent, calling for further analysis.  Introducing the coupling strengths $g_c$ defined in (\ref{g_c}) and arranging them in the increasing order $g\equiv g_1<g_2<\ldots < g_M$, the exact probability density for the distribution of scaled widths $y_n=\pi\Gamma_n/\Delta>0$ reads as follows \cite{somm99}:
\begin{equation}\label{exacbeta1}
  P_M(y) = \frac{d^2}{dy^2}\mathcal{F}_M(y)\,,
\end{equation}
where the `potential' function $\mathcal{F}_M(y)$ is given by
\begin{equation}\label{exacbeta1a}
  \mathcal{F}_M(y) = \frac{1}{\pi} \int_1^{\infty} \frac{dp_2}{\sqrt{p_2^2-1}}
  \frac{\chi_2(p_2)\,e^{-\frac{y}{2}p_2}}{\prod_{c=1}^M\sqrt{|g_c-p_2|}}
  \, f_M(p_2,y)
\end{equation}
\begin{equation}\label{exacbeta1b}
  f_M = \int_1^{p_2} \frac{dp_1(p_2-p_1)}{\sqrt{p_1^2-1}}
  \frac{\chi_1(p_1)\,e^{-\frac{y}{2}p_1}}{\prod_{c=1}^M\sqrt{|g_c-p_1|}}
  \,\phi_M(p_1,p_2,y)
\end{equation}
\begin{equation}\label{exacbeta1c}
  \phi_M(p_1,p_2,y) = \int_{-1}^{1} \frac{d\lambda}{2} (1-\lambda^2)
  \frac{e^{y\lambda}\prod_{c=1}^M(g_c-\lambda)}{(\lambda-p_1)^2(\lambda-p_2)^2}\,,
\end{equation}
with the factors $\chi_1(p)$ and $\chi_2(p)$ each showing a rather complicated alternating pattern of values $\pm 1$ or $0$ for various values of $p$, see \cite{somm99}. For the goals of our discussion it is enough to know only that $\chi_1(p)=1$ for $1\le p\le g$, whereas $\chi_2(p)=0$ in the same range. This distribution satisfies the Moldauer-Simonius relation exactly, yielding
\begin{equation}\label{MS}
  \aver{y}_\mathrm{MS} = \frac{1}{2}\sum_{c=1}^M\ln\left(\frac{g_c+1}{g_c-1}\right)
\end{equation}
for the average width. Note also that the validity of the exact distribution (\ref{exacbeta1}) at $M=1$ was verified in high-precision scattering experiments with microwave billiards \cite{kuhl08}.

\textbf{Weak coupling approximation.}\,-- Our goal now is to investigate the weak coupling limit $g=g_1\gg1$ of the exact expression considering, however, arbitrary values of the widths $y$. When performing such a limit we use that due to the nature of $\chi_{1,2}(p)$ factors the leading contribution to the integrals comes from the domain $(\lambda,p_1)\sim 1\ll g<p_2$. Correspondingly, we can replace factors $(\lambda-p_2)^2\approx p_2^2$, $p_2-p_1\approx p_2$, and $\sqrt{p_2^2-1}\approx p_2$ as well as approximate $g_c-p_1\approx g_c$ and $g_c-\lambda\approx g_c$ for all $c=1,\ldots M$.  Further analysis also shows that one can replace the upper limit $p_2$ in the integral (\ref{exacbeta1b}) by infinity. In this way we arrive at
\begin{equation}\label{F_approx}
  \mathcal{F}_M(y) = \Phi(y)\frac{2}{\pi}\int_1^{\infty}\frac{dp_2}{p_2^2}
  \frac{\chi_2(p_2)\,e^{-\frac{y}{2}p_2}}{\prod_{c=1}^M|1-g_c^{-1}p_2|^{1/2}}\,,
\end{equation}
where
\begin{equation}\label{phi1}
  \Phi(y) = \int_{-1}^{1}\frac{d\lambda}{4}(1-\lambda^2)
  e^{y\lambda}\int_1^\infty\frac{dp_1}{ \sqrt{p_1^2-1}} \frac{e^{-\frac{y}{2}p_1}}{(\lambda-p_1)^2}\,.
\end{equation}
After somewhat lengthy calculations one is able to evaluate the double integral (\ref{phi1}) in a closed form, finding
\begin{equation}\label{phi2}
 \Phi(y) = \frac{1}{2} \Bigl[ K_0\Bigl(\frac{y}{2}\Bigr) \Bigl(\cosh{y}-\frac{\sinh{y}}{y}\Bigr)
  + K_1\Bigl(\frac{y}{2}\Bigr) \sinh{y} \Bigl]\,,
\end{equation}
where $K_{0,1}(x)$ are the modified Bessel functions.

Noticing that $\Phi(y)$ in (\ref{F_approx}) is a $g$--independent factor and taking into account also that $p_2>g$ due to $\chi_2(p_2)$ function in the integrand, one may naively conclude that for $g\gg 1$ the leading order approximation to the width distribution in (\ref{exacbeta1}) is obtained by twice differentiating the $g$--dependent integral factor in (\ref{F_approx}). This results in
\begin{equation}\label{exacbeta1non-norm}
  P_{M}^{(\mathrm{naive})}(y) = \Phi(y)\,\mathcal{P}_{\chi^2}(y) \,,
\end{equation}
where
\begin{equation}\label{PTchi2}
  \mathcal{P}_{\chi^2}(y)=\int_1^{\infty} \frac{dp}{2\pi}
  \frac{\chi_2(p)\,e^{-\frac{y}{2}p}}{\prod_{c=1}^M|1-g_c^{-1} p|^{1/2}}\,.
\end{equation}
Note that expression (\ref{PTchi2}) has a specific meaning: it is just a version of the probability density of the $\chi^2$ distribution for weak non-equivalent channels \cite{somm99,alha95}, which can be shown to reduce to the standard form (\ref{PTdis}) for all the same $g_c=g$, with $g\approx\frac{1}{2\gamma}$ at $\gamma\ll 1$.  Therefore, all the nontrivial leading corrections to the PTD in such an approximation seem to be accumulated in the factor $\Phi(y)$. However the following problem arises: keeping only such leading term results in losing the correct normalization, being thus an inconsistent approximation to the probability density. Thus the other (higher-order) terms need to be also  considered, making the analysis more involved.

A resolution to that problem can be found by analysing and exploiting the factorised structure of (\ref{F_approx}) further. That form suggests an Ansatz providing the leading-order approximation to the width distribution (\ref{exacbeta1}) by differentiating the $g$--dependent factor only once as follows:
\begin{equation}\label{exacbeta1correct}
  \mathcal{P}_{M}(y) = -\frac{d}{dy} \left[ \Phi(y)\,\mathcal{N}_{M}(y)\right]\,.
\end{equation}
Function $\mathcal{N}_{M}(y)$ has the meaning of the cumulative (tail) distribution function associated with the density (\ref{PTchi2}), being normalized as $\mathcal{N}_{M}(0)=1$ and explicitly given by
\begin{equation}\label{Nchi2}
  \mathcal{N}_{M}(y) = \int_1^{\infty} \frac{dp}{\pi p}
  \frac{\chi_2(p)\,e^{-\frac{y}{2}p}}{\prod_{c=1}^M|1-g_c^{-1}p|^{1/2}}\,.
\end{equation}
Note that the form (\ref{exacbeta1correct}) satisfies the normalisation condition automatically. To ensure a positive probability density, the function $\Phi(y)\mathcal{N}_{M}(y)$ needs however to be monotonic. We can show that this is indeed the case for, e.g., $M=1,2$ where our Ansatz is therefore fully consistent\footnote{For $M\ge4$ the positivity of the ensuing density turns out to be mildly violated for a relatively small region of narrow widths. It can possibly be interpreted as a precursor of the depletion of narrow resonances due to the gap formation at larger $M$, see \cite{lehm95a}. Formally, it indicates on the noncommutative limits $M\gg1$ and $g\gg1$, thus nonuniform convergence due to the approximations involved.}. In particular, at $M=1$ we arrive after the straightforward integration to an attractively simple formula
\begin{equation}\label{singlechannel}
  \mathcal{P}_{M=1}(y) = -\frac{d}{dy}\left[\Phi(y)\,
 \mathrm{erfc}\left(\sqrt{\frac{g y}{2}}\right)\right]\,,
\end{equation}
where $\mathrm{erfc}(z)=1-\frac{2}{\sqrt{\pi}}\int_0^{z}e^{-t^2}dt$ is the complementary error function. The formulae (\ref{phi2}), (\ref{exacbeta1correct})--(\ref{singlechannel}) represent the consistent leading-order approximation of the resonance width distribution at weak coupling for any value of $y$, and as such constitute the main result of this paper.

\begin{figure}\label{fig1}
  \includegraphics[width=\columnwidth]{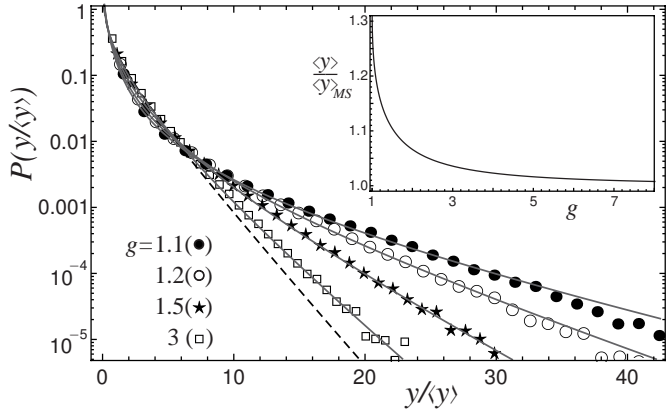}
  \caption{Distribution of the scaled resonance widths (in units of the mean width) for single-channel scattering at $g=1.1\,(\bullet)$, $1.2\,(\circ)$, $1.5\,(\star)$, and $3\,(\square)$. The solid lines stand for approximation (\ref{singlechannel}) whereas the dashed line is the Porter-Thomas expression ($M=1$).  The symbols correspond to numerics with $10^5$ realizations of 250 $\times$ 250 GOE random matrices (only 25 levels around the spectrum center were taken for each realization). The inset shows the comparison of the mean width $\aver{y}$ obtained with Eq.~(\ref{singlechannel}) to the exact (Moldauer-Simonius) expression $\aver{y}_{\mathrm{MS}}$ as a function of the coupling strength $g$.}
\end{figure}
It is instructive to analyse the single-channel case in more details. In the region of small widths, $y\ll1$, distribution (\ref{singlechannel}) coincides with the PTD up to the terms of the order $\sim\mathcal{O}(y^{3/2})$. It shows, however, significant deviations for large widths, with its leading asymptotic being
$\mathcal{P}_1(y\gg1) = e^{-(g-1)y/2} [ \frac{g-1}{2\sqrt{2g}}y^{-1}
  + \frac{4-(g-5)g}{8\sqrt{2g^3}}y^{-2} + \mathcal{O}(y^{-3})]$.
One sees that by formally setting $g=1$ here, we get the power-law behavior $\approx1/\sqrt{2}y^2$, which up to a constant reproduces the known exact one $\approx1/2y^2$. Correspondingly, the average width $\aver{y}$ computed by using (\ref{singlechannel}) develops $\frac{1}{\sqrt{2}}|\ln(g-1)|$ divergence at $g\to1$, in qualitative agreement with the exact expression $\aver{y}_{\mathrm{MS}}$ given by (\ref{MS}). This suggests that even though our formula (\ref{singlechannel}) has been derived under the assumption of $g\gg1$, it should properly approximate the exact distribution also at moderate $g\sim1$, provided that the rescaled widths $y/\aver{y}$ are used. To verify this procedure, we have performed numerical simulations with the GOE matrices and indeed found a good agreement, as is clearly demonstrated in fig.~1.

We briefly mention that similar findings hold at $M=2$ open channels (not shown here),  function (\ref{Nchi2}) being then given by $\mathcal{N}_2(y)=\sqrt{g_1g_2}\int_{g_1}^{g_2}\frac{dp}{\pi p}
[(g_2-p)(p-g_1)]^{-1/2}e^{-\frac{y}{2}p}$.

\textbf{Relation to spectral determinants.}\,-- To shed further light on the meaning and origin of the correction factor $\Phi(y)$, we note that it is independent of $M$ and $g$ and thus can be studied for the particular case $M=1$. We now reconsider this case by a different method inspired, in part, by our attempts to understand the nature of the heuristic proposal \cite{shched12}.  We start similarly with the exact joint probability density $\mathcal{P}\{\mathcal{E}_k\}$ of $N$ resonances in the complex plane, but find it slightly more convenient to work with the `fixed coupling' version of the model\cite{fyod11ox}:
\begin{equation}\label{jpd1a}
  \mathcal{P}\{\mathcal{E}_k\}\propto \gamma\prod_{n=1}^N S(\mathcal{E}_n)\prod_{m<n}
  \frac{|\mathcal{E}_m-\mathcal{E}_n|^2}{|\mathcal{E}_m-\mathcal{E}^*_n|}\,
  \delta\left(\gamma-\sum_{n=1}^N \frac{\Gamma_n}{2\lambda}\right)\,,
\end{equation}
where
$
 S(\mathcal{E}_n) = (\gamma\Gamma_n)^{-1/2}
 \exp[-\frac{N}{4\lambda^2}(\gamma^2\lambda^2+E_n^2-\Gamma_n^2/4)]\,
$
and we have omitted numerical constants which may depend on $N$ but are otherwise parameter-independent. Making use of this expression and following the method proposed in \cite{fyod99} for the $\beta=2$ case of the model (see \cite{fyod03r} for a detailed consideration), one can arrive in the limit $N\gg 1$ to a convenient representation for the ensuing density $\rho(E,\Gamma)$ of resonances around the point $\mathcal{E}=E-\frac{i}{2}\Gamma$. Note that at the spectrum center, $E=0$, this function has precisely the meaning of the corresponding resonance widths distribution. Under a weak condition $\Gamma\ll\lambda$ but otherwise for arbitrary resonance overlap $\frac{\Gamma}{\Delta}$, we find
\begin{equation}\label{PT1corrected}
  P_M(y)=\lim_{N\to\infty}\frac{\lambda}{N}\,\rho\Bigl(0,\frac{y\lambda}{N}\Bigr)\propto \frac{1}{\sqrt{y}}\,e^{-gy/2}\,\Psi_{\gamma}(y)\,,
\end{equation}
where $\Psi_{\gamma}(y)$ is the following spectral determinant ratio:
\begin{equation}\label{PT1factor}
  \Psi_{\gamma}(y) = \lim_{N\to\infty} C_N \left\langle \frac{|\det(\Heff^{(N-1)}+iy\lambda/2N)|^2}{|\det(\Heff^{(N-1)\dagger}+iy\lambda/2N)|}
  \right\rangle\,.
\end{equation}
Here the angular brackets stand for the GOE average over the Hermitian part of $\Heff^{(N-1)}$, which has now the reduced dimension of $N-1$. The constant $C_N$ is indicated for ensuring the limit is well defined. We stress that expressions (\ref{PT1corrected})--(\ref{PT1factor}) are the {\it exact representation} of the resonance density at $M=1$ in the limit considered, being valid for any values of width $y$ and at arbitrary coupling strength $g\geq1$. The above representation is especially convenient for our purposes, since it factorizes into a product of the PTD factor $\frac{1}{\sqrt{y}}\,e^{-gy/2}$ and the factor  $\Psi_{\gamma}(y)$, which therefore accumulates {\it all} corrections beyond the PTD.

Unfortunately, performing exactly the GOE average in (\ref{PT1factor}) at arbitrary coupling $\gamma\sim 1$ is a rather challenging task and remains an interesting outstanding problem of RMT. However, in the weak-coupling limit $\gamma\ll 1$ (i.e. $g\gg 1$) we can easily find a consistent approximation to that factor at the leading order in $\gamma$. This obviously amounts to simply setting $\gamma=0$ inside the RMT average, that is to neglecting the anti-Hermitian part by replacing $\Heff^{(N-1)}$ with $(N-1)\times(N-1)$ matrix $H$ from GOE in (\ref{PT1factor}). After doing this, we arrive at the problem of evaluating the following GOE average
\begin{equation}\label{factorRMT1}
  \lim_{\gamma=0}\Psi_{\gamma}(y) = \lim_{N\to\infty}C_N\left\langle\sqrt{\det{\left(H^2+y^2\lambda^2/4N^2\right)}}\right\rangle\,,
\end{equation}
where now $C_N=\frac{1}{\aver{\det|H|}}$. Actually, the right-hand side in (\ref{factorRMT1}) is similar to the objects appeared already in other RMT contexts, being the subject of a separate work \cite{FyodNock}. A particularly pleasant though not unexpected outcome of that work is that (\ref{factorRMT1}) can be evaluated in a closed form,  yielding $\lim_{\gamma=0}\Psi_{\gamma}(y)=\Phi(y)$ where $\Phi(y)$ is given exactly by (\ref{phi2}). This independently confirms the calculation done in the first part of this paper for $M=1$ by a completely different method and thus the internal consistency of all approximations made in the course of our derivation.

\textbf{Discussion and conclusions.}\,-- We are now in the position to  make contact to the proposal of ref.~\cite{shched12}. Basically, that proposal coincides with the form of the ``naive'' (not-normalized) expression (\ref{exacbeta1non-norm}) in which the correct factor $\Phi(y)$ was however replaced by a heuristic expression $\bigl(\frac{\sinh{y}}{y}\bigr)^{\beta/2}$.  Although such a form at $\beta=1$ shares similar large-$y$ asymptotic with $\Phi(y)$, it is manifestly different from the latter, rendering the proposal \cite{shched12}, in the strict sense, invalid. At this point it is appropriate to mention that in \cite{shched12} the authors actually arrived, though in a somewhat different way, to the problem of evaluating the same average featuring in the right-hand side of (\ref{factorRMT1}). However, they proposed to replace true GOE spectra there by a picket-fence model of equidistant levels. Though one may hope that the phenomenon of level repulsion could make it qualitatively sensible, such a replacement is certainly not a systematic approximation. In particular, neglecting spectral fluctuations is known to underestimate  interference effects related to non-orthogonal wavefunctions in open systems \cite{poli09b,savi13}. It is interesting to note, however, that similar manipulations with the exact GUE formulae for the resonance width distribution \cite{fyod96b} show that the ``true'' correction factor is actually given by  $\lim_{N\to\infty}\left\langle\det{\left(H^2+y^2\lambda^2/4N^2\right)}\right\rangle\propto \frac{\sinh{y}}{y}$, thus coinciding with the proposal \cite{shched12} at $\beta=2$. Such a coincidence appears nevertheless as a stroke of luck, though somewhat stronger level repulsion in the GUE case may indeed contribute to its success.

In conclusion, we used RMT framework to develop a systematic approximation to the resonance width distribution in weakly open quantum chaotic systems beyond the standard first-order perturbation theory. For the small number $M=1,2$ of open channels we found simple expressions which accurately reproduce the width distribution in the regime of moderately overlapping resonances, remaining qualitatively correct even at strong coupling to the continuum when resonances overlap considerably. As such, our results may be advantageous for practical applications. We also believe that the new representation (\ref{PT1corrected}) may prove to be useful for finding alternative derivation of the exact distribution in a simpler explicit form. Comparison with similar expressions at $\beta=2$ in \cite{fyod03r} makes us to anticipate that the factorized form (\ref{PT1corrected}) may retain its validity for $M>1$ after replacing the PTD factor with the density (\ref{PTchi2}). It remains an outstanding challenge to verify this from the first principles. Finally, it is worth mentioning that there is growing interest in this type of problems also in mathematical literature, see e.g. \cite{kill15}.

\acknowledgements
We are grateful to V.V. Sokolov for numerous discussions.  YVF acknowledges support by EPSRC grant EP/J002763/1 ``Insights into Disordered Landscapes via Random Matrix Theory and Statistical Mechanics''.


\end{document}